\documentclass[aps,pre,showpacs,floatfix]{revtex4}
\usepackage{graphicx}
\usepackage{times}
\usepackage{amsmath,amssymb}
\usepackage[latin1]{inputenc}
\usepackage{bm}
\def\be{\begin{equation}}
\def\ee{\end{equation}}

\begin{document}

\title{Survival probability of a particle in a sea of mobile traps: A tale of tails}

\author{S.B. Yuste$^1$, G.Oshanin$^2$, K.Lindenberg$^3$, O.B\'enichou$^2$ and J.Klafter$^4$}

\affiliation{
$^1$Departamento de F\'{\i}sica,  Universidad de Extremadura,
E-06071 Badajoz , Spain\\
$^2$Laboratoire de Physique Th{\'e}orique de la Mati\`ere
Condens\'ee, Universit{\'e} Pierre et Marie Curie, 4 Place Jussieu,
75252 Paris, France\\
$^3$Department of Chemistry and Biochemistry, University of
California, San Diego, La Jolla, CA 92093-0340 USA\\
$^4$School of Chemistry, Tel Aviv University, 69978 Tel
Aviv, Israel
}

\date{\today}

\begin{abstract}
We study the long-time tails of the survival probability $P(t)$ of
an $A$ particle diffusing in $d$-dimensional media in the presence of a concentration $\rho$ of
traps $B$ that move sub-diffusively, such that the mean square displacement of each trap grows as
$t^{\gamma}$ with $0\leq \gamma \leq 1$.
Starting from a continuous time random walk (CTRW) description of the motion of the particle and of
the traps, we derive lower and upper bounds for $P(t)$ and show that
for  $\gamma \leq 2/(d+2)$ these bounds coincide asymptotically, thus determining asymptotically exact
results.  The asymptotic decay law in this regime is exactly that obtained for immobile traps.  This
means that for sufficiently subdiffusive traps, the moving $A$ particle sees the traps as essentially immobile,
and Lifshitz or trapping tails remain unchanged.
For $\gamma > 2/(d+2)$ and $d\leq 2$ the upper and lower bounds again coincide, leading to a decay
law equal to that of a stationary particle. Thus, in this regime the moving traps see the particle as
essentially immobile. For $d>2$, however, the upper and lower bounds
in this $\gamma$ regime no longer coincide and the decay
law for the survival probability of the $A$ particle remains ambiguous.
\end{abstract}


\maketitle

\section{Introduction}

In the early 1960's, I.M. Lifshitz published his famous analysis
of the low-energy tails of the density $N(E)$ of states of an electron in a medium
with randomly scattered immobile impurities~\cite{lifa,lifb}.
He demonstrated that, in
$d$-dimensions, the spectrum $N(E)$ of the random Schr\"odinger operator
decays as
\begin{equation}
\label{dens}
N(E) \sim \exp\left(- const \;E^{-d/2}\right), \quad E \to 0.
\end{equation}
This exponential decay is in striking contrast to the polynomial decay associated with a periodic
Schr\"odinger operator, and is caused by the presence of arbitrarily large, albeit rare, regions
without impurities.

A decade later Balagurov and Vaks~\cite{bv} and Donsker and
Varadhan~\cite{dv} published their celebrated work on trapping kinetics in a medium with randomly
placed immobile traps (called $B$).  They found that the long-time decay of the
survival probability $P^B(t)$ of a particle (called $A$)
diffusing in such a $d$ dimensional medium
follows the stretched-exponential law
\begin{equation}
\label{tr}
P^B(t) \sim \exp\left(- k_d \rho^{2/(d+2)} (D_A t)^{d/(d+2)}\right),
\end{equation}
where $\rho$ is the mean density of traps, $D_A$ is the particle
diffusion coefficient, and $k_d$ is the $d$-dependent constant~\cite{dv}
\begin{equation}
\label{ll}
k_d = \left(\frac{d+2}{2 d}\right) \left(d v_d\right)^{2/(d+2)}\left(2
z_d^2\right)^{d/(d+2)}.
\end{equation}
Here $z_d$ is the first zero of the Bessel function
$J_{(d-2)/2}(z)$ and $v_d=2\pi^{d/2}/[d\,\Gamma(d/2)]$
denotes the volume of a $d$-dimensional sphere
of unit radius.  The superscript $B$ on the survival probability emphasizes that the traps $B$ are
immobile.  Balagurov and Vaks obtained the decay form in Eq.~(\ref{tr}) exactly for $d=1$.
While they furthermore deduced this behavior for general $d$ by noticing the
close mathematical connection between the trapping problem and the
Lifshitz problem, Donsker and Varadhan were the first to obtain the rigorous exact solution of
this essentially many-body problem in $d$ dimensions.
They determined not only the time dependence but also the decay
coefficient $k_d$ of Eq.~(\ref{ll}). Note that this law stands in stark contrast to
the purely exponential decay predicted by
standard chemical kinetics for the reaction $A+B\to B$ of mobile particles and traps, and even
to the predictions of the Smoluchowski approach based on a reaction-diffusion equation,
\begin{equation}
P(t)  \sim
\begin{cases}
\exp[- \rho (4 D t/\pi)^{1/2}],&  d=1\\
\exp[ - 4 \pi \rho D t/\ln(D t/a^2)],&  d=2 \\
\exp( - 4 \pi \rho D t), & d=3
\end{cases}
\end{equation}
where $D$ is the sum of the diffusion coefficients of the particle and the traps~\cite{kii}.

In mathematical analogy with the source of the Lifshitz tail, the decay law~(\ref{tr})
arises from the presence of arbitrarily large regions without traps in which the
particle can diffuse for a relatively long time before being trapped.
Although the survival probability of a particle in any one such
region is purely exponential, the ensemble average over a random distribution
of such trap-free voids produces the anomalously slow decay.

The decay law in Eq.~(\ref{tr}) has been also generalized
to trapping reactions on fractals and in inhomogeneous
structures~\cite{yossi}, to $A$ particles performing subdiffusive
motion~\cite{subdiffyuste1,subdiffyuste2,targetyuste,subdiffgleb}
or attached to the extremities of polymer chains
\cite{gleb2}, and also to agglomerated random distributions of
traps distributed on immobile polymer chains in
solution~\cite{gleb1} or in clusters~\cite{sasha}. The history of
this problem and many other results have been summarized in several
reviews (see, e.g.,~\cite{kehr,weiss}).

The survival probability Eq.~(\ref{tr}) [as well as the Lifshitz tail result of
Eq.~(\ref{dens})] is valid only when the traps (or impurities) are strictly immobile.
Indeed, if one allows them to
diffuse, no matter how small the diffusion coefficient, the particle
survival probability is described by a faster decay law.
As proved by Bramson and Lebowitz \cite{bl} (see also \cite{burl}), when both species
diffuse, the survival probability of the $A$ particle at long times is given by
\begin{equation}
\label{bramsleb}
P(t) \sim
\begin{cases}
\exp\left(- \rho c_1 t^{1/2}\right),&  d=1\\
\exp\left(- \rho c_2 t/\ln t \right), & d=2\\
\exp\left( - \rho c_3 t \right), & d=3.
\end{cases}
\end{equation}
The time dependences are the same as in the Smoluchowski problem, but the constants
$c_1, c_2$ and $c_3$ are in general different, may depend
on the diffusion coefficients of the particle and of the traps (but see below), and were not
determined in this original work.

When traps diffuse, the many-body trapping effects captured by
Eq.~\eqref{tr} are thus no longer applicable and the $A$ particle survival probability decays
with time according to the faster time dependencies in
Eq.~(\ref{bramsleb}). The underlying fluctuation mechanism governing the trapping dynamics has
changed when the traps are allowed to move. While
the time dependence in the decay laws~(\ref{bramsleb})  is
consistent with Smoluchowski-like results, which in fact represent
a two-body approximation, the decay amplitudes $c_d$
are not simply functions of the sum of the diffusion coefficients. In
particular, the exact form (including the coefficients) of the
leading large-$t$ behavior for $d=1$ and $d=2$ has only
recently been found by Bray and Blythe~\cite{bray} and,
surprisingly, $c_1$ and $ c_2$ depend only on the diffusion coefficient $D_B$ of the traps and
are independent of $D_A$, the diffusion coefficient of the $A$ particle!
This implies that the survival probability of the diffusing $A$ particle in this scenario is
asymptotically identical to that of an $A$ particle that remains still, that is, to $P^A(t)$.

This remarkable result has subsequently been extended to systems
in which both the particles and the traps move
subdiffusively~\cite{subdiffyuste1,subdiffyuste2,targetyuste,subdiffgleb}.
In particular, it is again found that the survival probability of a subdiffusive $A$ particle in a
sea of subdiffusive traps is identical to that of and $A$ particle that remains still (as in the
Bramson-Lebowitz scenario), $P(t) \sim P^A(t)$.
However, it is still not clear what happens when traps move subdiffusively and the particle is
diffusive.  In one dimension it was shown in~\cite{subdiffyuste1,subdiffyuste2} that the asymptotic survival
probability of the diffusing particle is the same when surrounded by subdiffusive traps
characterized by a mean square displacement that grows as $t^{\gamma}$ with $\gamma \leq 1$ as it is for
a stationary particle, \emph{provided the subdiffusive traps move sufficiently rapildy} ($\gamma>2/3$),
again akin to the Bramson-Lebowitz scenario.
However, few results seem to be currently available for $d>1$
and/or if the subdiffusive traps are ``too
slow."  An interesting question is then the following: will the trapping tails associated with
immobile traps withstand sufficiently slow
sub-diffusion, thus leading to a survival probability such as that of the Donsker-Varadhan result,
or will they switch to appropriately rescaled Bramson-Lebowitz forms?
This is the main question that we pose in this paper.

In this pursuit, we follow the general idea~\cite{bray}  of
constructing lower and upper bounds on the survival probability of a diffusive particle in a sea
of subdiffusive traps whose motion is described by a continuous time random walk that starts at time
$t=0$~\cite{ctrw}.   If these bounds coincide
asymptotically, then we can extract an exact asymptotic result for the survival probability of the particle.
We write the asymptotic survival probability in the form
\begin{equation}
\label{assume}
P(t)\sim
\begin{cases}
\exp\left(-\theta t^z\right), & d\neq 2\\
\exp\left(-\theta t^z/\ln t \right) & d=2,
\end{cases}
\end{equation}
where the constants $\theta$ and $z$ depend on $\gamma$ and on dimension, and explore whether a convergence of
upper and lower bounds provides the exponents $z$ and perhaps even the exponential prefactor
$\theta$.  We broadly anticipate our results by noting
that if the subdiffusive particles are sufficiently
slow, specifically if $\gamma \leq 2/(d+2)$, these bounds lead to a Donsker-Varadhan
behavior, and the Lifshitz or trapping tails thus remain unchanged.  If $\gamma > 2/(d+2)$
and $d \leq  2$, then the bounds lead to a Bramson-Lebowitz behavior
that is, a survival probability behavior associated with a stationary
particle (note that, albeit for a different system, this result is implicit in~\cite{subdiffgleb};
that work addresses diffusive particles in a fractal medium, while here we are considering
subdiffusive motion in Euclidean geometries). However, if $\gamma > 2/(d+2)$ and $d> 2$ we are not able to
establish a unique asymptotic behavior.  The detailed
results are exhibited later in the paper.

In Sec.~\ref{model} we formulate the model of moving particle and traps.  In Sec.~\ref{lowerbound}
we calculate a lower bound  on the survival probability of the particle, and in
Sec.~\ref{upperbound} we obtain two upper bounds. The consequences of these bounds on the survival
probability are collected in Sec.~\ref{collection}. A brief recapitulation of the results is given
in the concluding section.

\section{The Model}
\label{model}

In this section we formulate the general model that allows us to highlight
the approximations made to obtain upper and lower bounds on the survival probability.
Consider a $d$-dimensional system of volume $V$ containing a single
\emph{diffusive} $A$ particle of radius $a$
and $K$ randomly moving point traps $B$ (a finite trap radius would add nothing interesting
to the problem for our purposes).
The initial position of $A$ is the origin, and one realization of its trajectory is denoted by
${\bm a}_t$.  The starting points $B_0^{(k)},~k=1,2,\cdots K$ of the traps are randomly (Poisson)
distributed throughout the volume, and a trajectory of the $k$th trap is denoted by
${\bm B}_t^{(k)} = {\bm B}_0^{(k)} + {\bm b}_t^{(k)}$.

Next we define
\begin{equation}
\label{indicator}
v({\bm x}) =
\begin{cases}
\infty, &    |{\bm x}| \leq a, \\
 0, &    \text{otherwise}.
\end{cases}
\end{equation}
The indicator function $\Psi[{\bm a}_t,\{{\bm b}_t^{(k)}\}]$ of the event that the
$A$ particle has not met any of the $B$s up to time $t$ for a given
realization of their trajectories can then be written as
\begin{equation}
\Psi[{\bm a}_t,\{{\bm b}_t^{(k)}\}]=\prod_{k=1}^K  \Psi[{\bm a}_t,{\bm b}_t^{(k)}] = \prod_{k=1}^K \exp\left[- \int^t_0  v({\bm a}_{\tau} - {\bm
b}_{\tau}^{(k)} - {\bm B}_0^{(k)}) d\tau\right].
\end{equation}
Consequently the $A$ particle survival probability is
\begin{equation}
\label{l}
P(t)  =
E^{(a)}\left\{\prod_{k=1}^K \left( \frac{1}{V} \int_V d {\bm
B}_0^{(k)}  E^{(b)}\left\{  \Psi[{\bm a}_{t},{\bm b}_{t}^{(k)}] \right\}\right)\right\},
\end{equation}
where the symbol $E^{(a)}\left\{...\right\}$ denotes an average
over all $A$ particle trajectories ${\bm a}_t$ that start at the origin. The symbol
$E^{(b)}\left\{...\right\}$ denotes an average over all $B$
trajectories ${\bm b}_t$ whose starting point is $B_0$.  We have labeled each of the latter
trajectories by a trap label $k$, but since the traps move independently we can omit the
label.  Furthermore, we go to the thermodynamic limit
$K,V \to \infty$ while keeping the ratio $\rho = K/V$ fixed.
This leads to the $A$ survival probability
\begin{equation}
\label{3}
P(t)  =
E^{(a)}\Big\{e^{- \rho E^{(b)}\left\{W[{\bm a}_{t}-{\bm b}_{t}]\right\}}\Big\},
\end{equation}
where $W[{\bm a}_{t}-{\bm b}_{t}]$ is the functional of the trajectories ${\bm a}_{t}$
and ${\bm b}_{t}$
\begin{equation}
\label{W}
W[{\bm a}_{t}-{\bm b}_{t}] =
\int_{R^d}
d{\bm B}_{0} \left(1 - e^{- \int^t_0 v({\bm a}_{\tau} - {\bm b}_{\tau} -
{\bm B}_0) d\tau}\right).
\end{equation}
The exact problem has thus been reduced to an effective two-body problem
involving a single $A$ particle and a single $B$ trap. Nevertheless, it unfortunately does
not seem possible to evaluate the survival probability exactly from this expression,
the main mathematical difficulty being that the average over all possible
trajectories ${\bm b}_{t}$ has to be performed for each fixed ${\bm a}_{t}$.
Only after this average has been performed can one then go on to carry out the further average
over the $A$ particle trajectories. This appears to be a non-tractable
mathematical problem, and recourse has to be made to controllable
approximations. We do it here by constructing lower and upper bounds on $P(t)$ and identifying
conditions and regimes where these converge asymptotically.

\section{Lower Bound}
\label{lowerbound}

A lower bound was originally devised in~\cite{bray}
for diffusive particles and traps and was extended in~\cite{subdiffyuste1,subdiffyuste2,subdiffgleb}
to the case of (sub)diffusive particles
and traps. We adapt this method to the current situation in which the traps $B$ perform continuous-time
random walks, so that we may restate this bound in the language introduced above.
Following Bray and Blythe, we introduce a notional spherical volume $V_l$
of radius $l$ centered at the origin, and pick only those realizations of initial trap
distributions for which this volume is completely devoid of traps.  The probability that
the region will initially be empty of traps is
\begin{equation}
\label{void} P_{void}(l) \sim \exp\left(- v_d \rho l^d\right).
\end{equation}
We furthermore introduce the probability $P_A(l;t)$
that the $A$ particle does not leave the notional volume $V_l$
during time $t$~\cite{bray},
\begin{equation}
\label{mea}
P_A(l;t) \sim \exp\left(-z_d^2 D_A t/(l-a)^2\right).
\end{equation}
Thirdly, we introduce the probability that no $B$ particle enters the notional volume up to time
$t$, that is, the probability that an immobile $d$-dimensional target of radius $l$ survives up to
time $t$ in the presence of a concentration $\rho$ of traps all of which perform subdiffusive
motion~\cite{targetyuste}:
\begin{equation}
\label{ctrw-target}
P_B(l;t) \sim
\begin{cases}
\exp\left( - \rho \frac{\displaystyle (4\pi D_B t^{\gamma})^{d/2}}{\displaystyle
\Gamma(1-d/2)\Gamma(1+\gamma d/2)}\right), &   d < 2 \\
\exp\left( - \rho \frac{\displaystyle 4\pi D_B t^{\gamma}}{\displaystyle \Gamma(1+\gamma)
\ln\left(4D_B t^{\gamma}/l^2\right)} \right),&  d=2 \\
\exp\left( - \rho \frac{\displaystyle 2 (d-2)\pi^{d/2} l^{d-2} D_B t^{\gamma}}{\displaystyle
\Gamma(d/2)\Gamma(1+\gamma)}\right), & d> 2.
\end{cases}
\end{equation}
Here $\Gamma$ is the gamma function, and $D_B$ is the anomalous diffusion coefficient of the traps,
that is, the coefficient in the mean square displacement relation $\langle r^2 \rangle =
2dD_B t^\gamma/\Gamma(1+\gamma)$.

Since the functional in Eq.(\ref{l}) is positive definite, the two
latter constraints on the trajectories of $A$ and the $B$s
naturally lead to a lower bound on the
survival probability $P(t)$. Furthermore, for these constrained trajectories
the functional in Eq.(\ref{W}) is strictly equal
to zero, and hence Eq.~(\ref{l}) restricted in this way is equal to unity.
As a consequence, the probabilities associated with the
random processes ${\bm a}_{t}$ and ${\bm b}_{t}$ subject to these constraints
can simply be factored, immediately leading to the lower bound on $P(t)$
\begin{equation}
\label{lb}
P(t) \geq P_{void}(l) P_A(l;t) P_B(l;t) \equiv P_L(l;t).
\end{equation}
Finally, we note that this lower bound $P_L(l;t)$ is in fact a family of lower bounds dependent
on the radius $l$ of the notional volume separating particle and traps. This radius can be chosen
to give the best lower bound, that is, the maximal lower bound, which we simply denote as $P_L(t)$.
The optimal radii are shown explicitly in Appendix~\ref{appa}, and are shown to depend on
dimensionality and on $\gamma$.
The associated optimal lower bounds for the survival probability of $A$, which also depend on
dimensionality and on $\gamma$, then immediately
follow.  We thus have
\begin{equation}
\label{lbtight}
P(t) \geq P_L(t)
\end{equation}
where, for $d<2$,
\begin{equation}
\label{optlower1d}
P_L(t) \sim
\begin{cases}
\exp \left( -k(d) \rho^{2/(d+2)} (D_A t)^{d/(d+2)} \right), &\gamma < \frac{2}{d+2} \\
\exp \left( - \frac{ \displaystyle \rho (4\pi D_B t^\gamma )^{d/2}}{\displaystyle \Gamma(1-d/2)
\Gamma(1+\gamma d/2)}\right), & \gamma >
\frac{2}{d+2}.
\end{cases}
\end{equation}
For $d=2$ we have
\begin{equation}
\label{optlower2d}
P_L(t)\sim
\begin{cases}
\exp\left( - (4\pi z_2^2 \rho D_A t)^{1/2} \right), & \gamma < \frac{1}{2} \\
\exp \left( - \frac{\displaystyle 8\pi \rho D_B t^\gamma}{\displaystyle (2\gamma-1) \Gamma(1+\gamma)\ln t } \right), & \gamma
> \frac{1}{2}.
\end{cases}
\end{equation}
For $d>2$ the results are
\begin{equation}
\label{optlower3d}
P_L(t) \sim
\begin{cases}
\exp \left( -k(d) \rho^{2/(d+2)} (D_A t)^{d/(d+2)} \right), & \gamma < \frac{2}{d+2}  \\
\exp \left( - d \left( \frac{\displaystyle (d-2)\pi^{d/2} \rho D_B t^\gamma}{\displaystyle
\Gamma(d/2)\Gamma(1+\gamma)}\right) ^{2/d}
\left( \frac{\displaystyle z_d^2 D_A t}{\displaystyle d-2}\right)^{(d-2)/d}\right), &
\gamma > \frac{2}{d+2}.
\end{cases}
\end{equation}
There is thus a change in behavior of the lower bound in all dimensions when
$\gamma$ crosses the value $2/(d+2)$.

\section{Upper Bounds}
\label{upperbound}

\subsection{Pascal Principle}
\label{pascal}

An upper bound
on the survival probability $P(t)$ of the diffusive particle was recently
derived on the basis of the so-called ``Pascal Principle," which states that the $A$ particle survives
longer if it remains still than if it moves. The problem with a static $A$ particle and moving traps
is the so-called ``target problem," and consequently we label this upper bound on the survival
probability as $P_{U,target}(t)$.  The Pascal Principle was conjectured in~\cite{bray}
and was  proved in~\cite{pascal2} for $d<2$ for conventional diffusive motion.  In~\cite{pascal1}
the statement was proved for a rather general class of random walks on $d$-dimensional lattices.
A similar statement was introduced more than a decade earlier in~\cite{pascal3} in the context of
excitation energy migration.  This upper bound is given by Eq.~(\ref{ctrw-target}) obtained
in~\cite{targetyuste} if we set $l=a$, the radius of the $A$ particle.  For visual ease
we explicitly rewrite Eq.~(\ref{ctrw-target}) with this replacement,
\begin{equation}
\label{upper1}
P_{U,target}(t) \sim
\begin{cases}
\exp\left( - \rho \frac{\displaystyle (4\pi D_B t^{\gamma})^{d/2}}{\displaystyle
\Gamma(1-d/2)\Gamma(1+\gamma d/2)}\right), &   d < 2 \\
\exp\left( - \rho \frac{\displaystyle 4\pi D_B t^{\gamma}}{\displaystyle \Gamma(1+\gamma)
\ln\left(4D_B t^{\gamma}/a^2\right)} \right),&  d=2 \\
\exp\left( - \rho \frac{\displaystyle 2\pi^{d/2} a^{d-2} D_B t^{\gamma}}{\displaystyle
\Gamma(d/2)\Gamma(1+\gamma)}\right), & d> 2.
\end{cases}
\end{equation}
To clearly tie together the various notations introduced so far, we note that
\begin{equation}
P_B(a;t) \equiv P^A(t) \equiv P_{U,target}(t)
\end{equation}
and
\begin{equation}
\label{uppertarget}
P(t) \leq P_{U,target}(t)
\end{equation}

However, this upper bound when associated with the lower bound is not always sufficiently tight to
provide the desired information about the asymptotic survival probability of the diffusive particle.
We thus introduce and alternative new upper bound, which in some cases is lower than the above.  In
the next section we then explicitly pick the bounds to be used and exhibit the information that can
be obtained from them.

\subsection{Anti-Pascal Principle}
\label{antipascal}

This new upper bound
is based on what we will call the
``Anti-Pascal Principle."  We will show that
the worst possible strategy for traps in their search for a target is to remain immobile.
Random motion, even uncorrelated with the motion of the target, enhances the probability to encounter the
target.  In other words, the diffusing particle $A$ survives longer if the traps remain still.
The problem with a moving $A$ particle in a sea of static traps is the so-called ``trapping
problem," and so we label this upper bound on the survival probability as
$P_{U,trapping}$. Again, to clearly tie together various notations we note that
$P^B(t) \equiv P_{U,trapping}(t)$, and the upper bound just introduced then says that
\begin{equation}
\label{imm}
P(t) \leq P_{U,trapping}(t) = P^B(t) = E^{(a)}\Big\{e^{- \rho W[{\bm a}_{t}]}\Big\}.
\end{equation}

To prove Eq.~\eqref{imm} we make use of Jensen's inequality for convex
functions, which for our model can be stated as
\begin{equation}
e^{- \rho E^{(b)}\Big\{W[{\bm a}_{t} - {\bm b}_{t}]\Big\}} \leq
E^{(b)}\Big\{ e^{- \rho W[{\bm a}_{t} - {\bm b}_{t}]} \Big\}
\end{equation}
(Note that this inequality is generally used to derive a \emph{lower}
bound. Indeed, when applied to the $E^{(a)}$ average in
Eq.~(\ref{3}) it yields a lower bound which is exactly the
Smoluchowski-type result~\cite{burl}.)
Consequently, we have the following upper bound:
\begin{equation}
\label{4}
P(t) \leq  E^{(a)}\left\{E^{(b)}\left\{ e^{- \rho  W[{\bm a}_t - {\bm b}_t]}
\right\} \right\} \equiv
E^{(c)}\left\{e^{- \rho W[{\bm c}_t]}     \right\},
\end{equation}
where $E^{(c)}\left\{...\right\}$ denotes an average with respect
to the trajectories ${\bm c}_t = {\bm a}_t - {\bm b}_t$ of a
``fictitious" particle $C$ of radius $a$ which starts its motion at ${\bm B}_0$.

Note that $W[{\bm c}_t]$ has a clear geometric interpretaion - it defines the volume swept by the
fictitious particle $C$ within the time interval $(0,t)$ and is thus an analog of the so-called
Wiener sausage for conventional diffusive motion.  Clearly, $W[{\bm c}_t]$ is a \emph{non
decreasing} function of time $t$ or, more precisely, of the number $N$ of jumps made by the
fictitious particle within the time interval $(0,t)$.  This number has two contributions,
$N=N_A+N_B$, where $N_A$ is the number of jumps made by particle $A$ and $N_B$ the number of jumps
made by a trap. Setting $N_B=0$, i.e., supposing that the trap is immobile, clearly diminishes the
total number of jumps and hence diminishes $W[{\bm c}_t]$. In this way we tighten the bound in
Eq.(\ref{4}) to arrive at the desired inequality~(\ref{imm}) 
We have thus established the second upper bound (\ref{imm}) for the survival
probability of $A$.  The maximal information about the survival probability is thus obtained from
the lower bounds presented in Eqs.~(\ref{lbtight})-(\ref{optlower3d}) and the smaller of the upper
bounds~(\ref{uppertarget}) with~(\ref{upper1}) and~(\ref{imm}) with~(\ref{tr}).

\section{Collecting Results and Bounding the Survival Probability}
\label{collection}

\begin{figure}[b]
\def\capfrac{1}
\includegraphics[width=.75\textwidth]{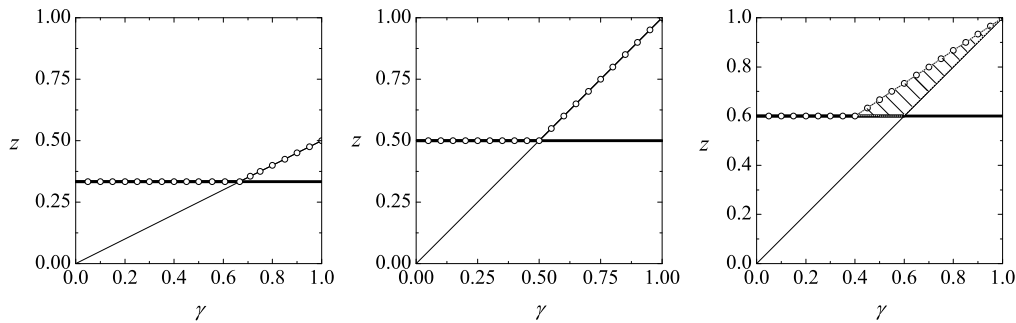}
\caption{
Asymptotic results for the survival probability of a diffusive particle in a sea of
subdiffusive traps.  Plotted is the exponent $z$ as expressed in Eq.~(\ref{assume}),
or bounds on this exponent, 
as a function of the trap subdiffusion exponent $\gamma$.  First panel: $d=1$; second panel: $d=2$.
Third panel: $d=3$. 
The solid lines are upper bounds; specifically, the thick solid lines are Pascal upper bounds (immobile
particle) while the thin solid lines are Anti-Pascal upper bounds (immobile traps). The solid lines
with superimposed circles are lower bounds.  Thus the first and second panels show that for $d\leq
2$ the Pascal (Donsker-Varadhan) and lower bound lines coincide for $\gamma \leq 2/(d+2)$.
The behavior changes when $\gamma$
crosses the value $2/(d+2)$, leading to a coincidence of the Anti-Pascal (Bramson-Lebowitz) and
lower bound lines. For $d\leq2$ the asymptotic exponent is thus obtained for all values of $\gamma$.
The third panel for $d=3$  shows convergence of the upper (Pascal) and lower bounds for $\gamma\leq
2/(d+2)$. However, when $\gamma > 2/(d+2)= 2/5$ the situation is uncertain.  The
exponent $z$ now lies in the triangular region delimited by the lower bound and the Pascal ($2/(d+2)
\leq \gamma \leq d/(d+2)$) or Anti-Pascal ($d/(d+2) \leq \gamma \leq 1$) upper bound.
}
\label{fig1}
\end{figure}

In this section we collect our detailed results, exhibited by displaying values for the constants
$z$ and $\theta$ in Eq.~(\ref{assume}) when the upper and lower bounds converge asymptotically, and providing
bounds for the exponent $z$ when they do not.  The results for the exponent $z$ as a function of the
subdiffusive trap exponent $\gamma$ are shown for integer dimensions $d=1$, $2$, and $3$ in
Fig.~\ref{fig1}.
Even before displaying the explicit results we note the following:
\begin{itemize}
\item
When $\gamma<2/(d+2)$ the asymptotic survival probability of the diffusive $A$ particle in the sea of
subdiffusive traps is the same as it is for \emph{immobile traps}.  The question asked in the introduction,
whether trapping tails associated with immobile traps withstand sufficiently slow sub-diffusion, is
thus answered in the affirmative, with a precise dimension-dependent characterization of what
is ``sufficiently slow."
This behavior, obtained by the asymptotic convergence of the bounds $P_L(t)$ and $P_{U,trapping}(t)$, is
shown for the exponent $z$ by the thick solid lines with superimposed circles
in Fig.~\ref{fig1}.  In this regime the survival probability is thus of the Donsker-Varadhan form.
\item
When $\gamma$ crosses the value $2/(d+2)$ there is a kind of dynamical phase transition. For $d\leq
2$ the survival probability of the diffusive particle in the sea of now more rapidly moving
subdiffusive traps decays as it would if the \emph{particle remains immobile}. This is indicated by
the thin solid lines with superimposed circles in the first two panels of Fig.~\ref{fig1}, and
results from the asymptotic convergence of the bounds
$P_L(t)$ and $P_{U,target}(t)$.  In this regime the survival probability is thus of the
Bramson-Lebowitz form generalized to subdiffusive traps.
\item
For $d>2$ and $\gamma> 2/(d+2)$ the situation is left somewhat uncertain: we are only able to bound
the decay exponent but not determine it uniquely, because neither upper bound converges
asymptotically to the lower bound. All we can say is that the survival probability decay exponent
$z$ lies in the triangular region bounded by the thin solid line with superimposed circles
(associated with the lower bound), the
thick solid line (associated with $P_{U,trapping}$), and the thin solid line (associated with
$P_{U,target}$), as indicated in the figure caption.
\end{itemize}

\begin{table}
\begin{center}
\begin{tabular}{ | c | c | c | c | c | }
\hline
Dimension $d$ & Trap Subdiffusive & Optimal & Survival Probability & Survival Probability \\
	      & Exponent $\gamma$ & $~$ Upper Bound$~$ & Exponent $z$ & Prefactor $\theta$ \\
\hline
&&&&\\
$~~~1\leq d < 2 ~~~$ & $~~~0\leq \gamma \leq \frac{\displaystyle 2}{\displaystyle d+2}~~~$ &
$P_{U,trapping}$&
$~~~\frac{\displaystyle
d}{\displaystyle d+2}~~~$ & $~~~k(d) \rho^{2/(d+2)} D_A^{d/(d+2)}~~~$ \\
&&&&\\
& $~~~ \frac{\displaystyle 2}{\displaystyle d+2} \leq \gamma \leq 1 ~~~$ &
$P_{U,target}$ &
$~~~\frac{\displaystyle \gamma d}{\displaystyle 2} ~~~$&
$~~~ \frac{\displaystyle \rho(4\pi D_B)^{d/2}}
{\displaystyle \Gamma(1-d)\Gamma(1+\gamma d/2)} $\\
&&&&\\
\hline
&&&&\\
$~~~ d=2 ~~~$ & $~~~0\leq \gamma \leq \frac{\displaystyle 1}{\displaystyle 2}~~~$ &
$P_{U,trapping}$&
$~~~\frac{\displaystyle 1}{\displaystyle 2} ~~~ $& $~~~ k(2) (\rho D_A)^{1/2} ~~~$ \\
&&&&\\
& $~~~ \frac{\displaystyle 1}{\displaystyle 2} < \gamma \leq 1 ~~~$ &
$P_{U,target}$ &
$~~~ \gamma ~~~$ &
$~~~\frac{\displaystyle 1}{\displaystyle \gamma} \leq \frac{\displaystyle
\Gamma(1+\gamma)}{\displaystyle 4\pi \rho D_B} \theta \leq \frac{\displaystyle 2}{\displaystyle
2\gamma - 1} ~~~$ \\
&&&&\\
\hline
&&&&\\
$~~~ d>2 ~~~$ & $~~~0\leq \gamma \leq \frac{\displaystyle 2}{\displaystyle d+2}~~~$ &
$P_{U,trapping}$&
$~~~\frac{\displaystyle d}{\displaystyle d+2}~~~$  & $~~~k(d) \rho^{2/(d+2)} D_A^{d/(d+2)}~~~$ \\ 
&&&&\\
& $~~~\frac{\displaystyle 2}{\displaystyle d+2} \leq \gamma \leq \frac{\displaystyle d}{\displaystyle
d+2}~~~$ &
$P_{U,target}$ &
$~~~ \frac{\displaystyle d}{\displaystyle d+2} \leq z \leq \frac{\displaystyle
d-2}{\displaystyle d} + \frac{\displaystyle 2\gamma}{\displaystyle d} ~~~$& $~~~$ ? $~~~$ \\
&&&&\\
& $~~~\frac{\displaystyle d}{\displaystyle d+2} \leq \gamma \leq 1 ~~~$ &
$P_{U,target}$ &
$~~~\gamma \leq z \leq
\frac{\displaystyle d-2}{\displaystyle d} + \frac{\displaystyle 2\gamma}{\displaystyle d} ~~~$& $~~~$
? $~~~$ \\
&&&&\\ 
\hline
\end{tabular}
\end{center}
\label{table}
\caption{Collected results for the asymptotic survival probability exponent $z$ and prefactor $\theta$.}
\end{table}

Finally, we collect the explicit results described above in Table~\ref{table}. 
The reported results are for the exponents $z$ and the prefactors $\theta$ in
Eq.~(\ref{assume}).  Again, we note that the results for $z$ are sketched explicitly for
$d=1,2,3$ in Fig.~\ref{fig1}.

\section{Conclusions}
\label{conclusions}

In conclusion, we have rigorously determined the survival probability of a particle
diffusing in a $d$-dimensional medium in the presence of a
concentration of traps performing subdiffusive random motion.
We have arrived at our results by calculating a lower bound and two alternative upper bounds for the
survival probability. One of the upper bounds, based on the ``Pascal Principle," is obtained by assuming
the particle to remain immobile.  The other, based on the ``Anti-Pascal Principle," is found by assuming
that the traps remain immobile.  We can then choose the tighter (lower) upper bound. Results for the
asymptotic survival probability of the particle can thus be extracted if the lower bound and one
of the upper bounds converge asymptotically.

Following this procedure, we have shown that when the dynamical exponent $\gamma$
characterizing the growth of the second moment of the displacement of the traps is less
than $2/(d+2)$, that is, if the traps move sufficiently slowly, the decay of the survival
probability of the diffusing particle in any dimension is given exactly by the Donsker-Varadhan result
obtained for immobile traps.  When the traps move more rapidly than this, i.e.,
when $\gamma > 2/(d+2)$, then in
dimensions $d\leq 2$ the survival probability of the particle is identical to that of a stationary
$A$ particle in a sea of mobile traps.  For higher dimensions, $d>2$, our results do not uniquely
determine the survival probability of the particle in this $\gamma$ regime, but they do provide
tighter bounds than previously known.  We close by noting the well-known difficulties that may
stand in the way of reaching and verifying some asymptotic results via numerical simulations
(see~\cite{subdiffyuste2} and references therein). Such simulations continue to pose an interesting
challenge.

\section*{Acknowledgments}

The research of S.B.Y. has been supported by the Ministerio de Educaci\'on y Ciencia (Spain) through grant
No. FIS2007-60977 (partially financed by FEDER funds). K.L. gratefully acknowledges the support of the National
Science Foundation under Grant No. PHY-0354937. The research of O.B. and G.O. is partially supported
by Agence Nationale de la Recherche (ANR) under grant ``DYOPTRI - Dynamique et
Optimisation des Processus de Transport Intermittents''.

\appendix
\section{Optimal Notional Volume}
\label{appa}

In this appendix we explicitly show the optimal radius $l$ that leads to the maximal lower bound on
the survival probability of the diffusive particle in the presence of a sea of subdiffusive point
traps.  Combining the expressions in Eqs.~(\ref{void}), (\ref{mea}) and
(\ref{ctrw-target}) and differentiating their product with
respect to $l$, we find that the optimal $l$ depends on dimensionality and on $\gamma$.
For $d<2$ we find
\begin{equation}
\label{optl1d}
l\sim \left( \frac{2z_d^2 D_A t}{dv_d\rho}\right)^{1/(d+2)}.
\end{equation}
For $d=2$ we obtain
\begin{equation}
\label{optl2d}
l \sim
\begin{cases}
\left( \dfrac{\displaystyle z_2^2 D_A t}{\displaystyle \pi \rho}\right) ^{1/4}, &\gamma < \frac{1}{2}\\
\left( \dfrac{\displaystyle z_2^2 \Gamma(1+\gamma) D_A t}{\displaystyle 4\pi  \rho D_B t^{\gamma}}\right)^{1/2}
\ln\left(\dfrac{ 16\pi \rho D_B^2 t^{2\gamma}}{ z_2^2 \Gamma(1+\gamma) D_A t} \right), & \gamma>\frac{1}{2}.
\end{cases}
\end{equation}
For $d>2$ we find
\begin{equation}
\label{optl3d}
l\sim
\begin{cases}
\left( \frac{\displaystyle 2z_d^2 D_A t}{\displaystyle d v_d \rho}\right) ^ {1/(d+2)}, & \gamma < \frac{2}{d+2}\\
\left( \frac{\displaystyle z_d^2 \Gamma(d/2)\Gamma(1+\gamma)
D_A t^{1-\gamma}}{\displaystyle (d-2)^2 \pi^{d/2}\rho D_B}\right)^{1/d}, & \gamma >
\frac{2}{d+2}.
\end{cases}
\end{equation}

\end{document}